\documentclass[twocolumn,showpacs,amsmath,amssymb,letterpaper,pre]{revtex4}

\usepackage{graphicx}
\usepackage{dcolumn}
\usepackage{bm}

\providecommand\bnabla{\boldsymbol{\nabla}}

\newcommand{\vv}[1]{\boldsymbol{#1}}
\def\acos{\mathop{\rm acos}\nolimits}

\newcommand{\rz}{R_0}

\newcommand{\rmax}{R_\text{max}}
\newcommand{\rres}{R_\text{res}}
\newcommand{\rmoins}{R_{-}}
\newcommand{\rplus}{R_{+}}
\newcommand{\vplus}{V_{+}}
\newcommand{\vmoins}{V_{-}}
\newcommand{\iplus}{I_{+}}
\newcommand{\imoins}{I_{-}}
\newcommand{\xp}{x_{+}}
\newcommand{\yp}{y_{+}}
\newcommand{\dx}{\Delta x}
\newcommand{\dsurd}[2]{\frac{\partial #1}{\partial #2}}
\newcommand{\sdsurd}[2]{\partial #1/\partial #2}

\newcommand{\xmun}{x_{m_1}}
\newcommand{\rc}{R_0^c}
\newcommand{\lhsrayleighp}{R\ddot{R}+ \troisdemi \dot{R}^2}
\newcommand{\troisdemi}{\frac{3}{2}}

\newcommand{\asd}{\frac{A}{2}}
\newcommand{\pact}{{\mathcal P}}
\newcommand{\pac}{P}
\newcommand{\ri}{x_i}
\newcommand{\pgradt}{\dsurd{\pact}{\ri}}
\newcommand{\pgrad}{G_i}
\newcommand{\fbun}{{\vv{F_{B}}}}

\newcommand{\grad}{\vv{\nabla}}
\newcommand{\vvr}{\vv{r}}

\begin{document}


\title{Analytical expressions for primary Bjerknes force on inertial
  cavitation bubbles.}

\author{Olivier Louisnard}
\email{louisnar@enstimac.fr}

\affiliation{%
  RAPSODEE research center, UMR EMAC-CNRS 2392, Ecole des Mines
  d'Albi, 81013 Albi, France}%


\date{\today}

\begin{abstract}
  The primary Bjerknes force is responsible for the quick
  translational motion of radially oscillating bubbles in a sound
  field.  The problem is classical in the case of small-amplitude
  oscillations, for which an analytical expression of the force can be
  easily obtained, and predicts attraction of sub-resonant bubbles by
  pressure antinodes. But for high-amplitude sound fields, the bubbles
  undergo large amplitude nonlinear oscillations, so that no
  analytical expression of the force is available in this case.  The
  bubble dynamics is approximated on physical grounds, following the
  method of \citeauthor{hilgenbrennergrosslohse98} [J. Fluid Mech.,
  \textbf{365}, 171 (1998)], but carefully accounting for surface
  tension. The analytical expression of the maximum radius of the
  bubble is recovered, the time of maximum expansion is noticeably
  refined, and an estimation of the collapse-time is found.  An
  analytical expression for the time-varying bubble volume is deduced,
  and the Bjerknes force is obtained in closed form. The result is
  valid for any shape of the sound field, including purely standing or
  purely traveling waves, and is ready to use in a theoretical model
  of bubble clouds evolution.  Besides, the well-known sign inversion
  of the Bjerknes force for large standing waves is recovered and the
  inversion threshold in the parameter space is obtained analytically.
  The results are in good agreement with numerical simulation and
  allow a quantitative assessment of the physical parameters effect.
  It is found that either reducing surface tension, or increasing the
  static pressure, should produce a widening of the bubble-free region
  near high-amplitude pressure antinodes.
\end{abstract}

\maketitle

\section{Introduction}
When excited by a sinusoidal sound field, gas bubbles undergo radial
oscillations. Most of the practical applications of this phenomenon,
known as acoustic cavitation, use high-amplitude sound fields, of
typical amplitude greater than the static pressure, so that the liquid
is under tension for some part of the cycle.  In such conditions,
whatever the frequency, two distinct dynamic bubble behaviors can be
clearly divided by the so-called Blake threshold
\cite{blake49rectif,neppirasphysrep,akhatovgumerov97,%
  hilgenbrennergrosslohse98}: in the tension phase, very small bubbles
are retained to grow by surface tension. Conversely, larger ones
suffer an explosive expansion followed by a violent collapse,
responsible for chemical \cite{suslicknato}, mechanical effects
\cite{lauterbornreview99,krefting2004} and sonoluminescence
\cite{gaitancrum92,putterman2000,brenner2002}. The latter oscillation
regime is known as ``inertial cavitation''.

Bubbles in liquids experience various hydrodynamic forces. The
buoyancy force is the most familiar one, and is the pressure force
that an sphere of liquid replacing the bubble would experience. This
remains true in a accelerating liquid \cite{magnaudetasme97}, and the
generalized buoyancy force experienced by the bubble is $-V\bnabla
\pact$ where $\pact (\vvr,t)$ is the pressure that would exist at the
center of the bubble if it were absent, and $V(t)$ the bubble volume.
For a bubble oscillating radially in a sound field, both
$\pact(\vvr,t)$ and $V(t)$ are oscillatory quantities so that the
time-average of the product over one cycle is not zero.  The bubbles
experiences therefore a net force known as ``primary Bjerknes force''
\cite{bjerknes,blake49}:
\newcommand{\fbjun}{\vv{F_{B_1}}}
\begin{equation}
  \label{defbjerknesintro}
  \fbun = - \left<V\bnabla \pact \right>
\end{equation}
The Bjerknes force can be easily calculated from the knowledge of both
the shape of the sound field and the bubble dynamics. A classical
result is that for low-amplitude standing waves, sub-resonant bubbles
are attracted by pressure antinodes, while bubbles larger than
resonant size are repelled
\cite{goldmanringo,crumeller70,leightonwalton}. For the case of strong
driving pressures, sub-resonant inertial bubbles can also be attracted
by pressure antinodes, which constitutes the basic principle of SBSL
levitation cells \cite{gaitancrum92,barber}. However it has been shown
by numerical calculations that above a given threshold, the primary
Bjerknes force on sub-resonant inertial bubbles undergoes a sign
change \cite{akhatov97une}. This behavior is due to the resonance-like
response curve (termed as ``giant resonance'' by Lauterborn and
co-workers \cite{responsecurvesandmore}) of the bubble just above the
Blake threshold, which is a physical consequence of the effect of
surface tension. Experiments indeed demonstrate that above a certain
driving level, no bubbles are visible in the neighborhood of large
pressure antinodes \cite{mettin99}. %

Quantitative agreement between theory and experiment has been found in
the case of linear or quasi-linear oscillations \cite{crumeller70}.
Particle simulations \cite{mettin99,parlitz99} were also found in
excellent agreement with recent experiments involving inertial bubbles
\cite{kochkrefting2003}.  While the Bjerknes force can be calculated
analytically for linear bubble oscillations, only numerical results
can yet be found for inertial bubbles \cite{akhatov97une,mettin2007}.
An analytical expression for the latter would first be helpful in
particle or continuum models, describing the self-organization of
bubbles, in order to get more efficient calculations. Furthermore,
analytical results allow a direct assessment of the sensitivity of the
force to the physical parameters, and the establishment of scaling
laws.  These two objectives motivated this study.

Owing to the strong nonlinearity of the bubble dynamics equations,
inertial cavitation has long been thought intractable analytically, up
to the seminal papers of \citet{lofstedt93} and
\citet{hilgenbrennergrosslohse98}, who demonstrated that several terms
of the Rayleigh-Plesset equation (RP) could be neglected during the
explosive expansion of the bubble. This theoretical breakthrough
allowed to obtain scaling laws for the maximum radius of the bubble
and the time of maximum expansion.  In this paper, we closely follow
the approach of \citet{hilgenbrennergrosslohse98} and refine their
analytical solutions in order to account more precisely for the effect
of surface tension.  The approximate dynamics found are then used to
obtain an analytical expression of the bubble volume. The latter are
then conveniently recast in order to obtain the Bjerknes force
(\ref{defbjerknesintro}) in closed form, in any acoustic field,
including the two extreme cases of traveling and standing waves.
Finally, in the latter case, we seek an approximate expression of the
Bjerknes force inversion threshold, evidencing the role of surface tension.

\section{Primary Bjerknes force}
\label{secbjerknes}
\subsection{Acoustic field}
We assume that the acoustic field in the liquid is mono-harmonic at
angular frequency $\omega$, and defined in any point $\vvr$ by
\begin{equation}
  \label{defpaco}
  \pact (\vvr,t) = \pac (\vvr) \cos
  \left[\omega t+\phi(\vvr) \right].
\end{equation}
This expression may represent a traveling wave, a standing wave, or
any combination of both. We also define the pressure gradient in
general form as
\begin{equation}
  \label{defpgrad}
  \pgradt  (\vvr,t) = \pgrad (\vvr) \cos
  \left[\omega t + \psi_i(\vvr) \right],
\end{equation}
where the fields  $\pgrad$ and $\psi_i$ can be expressed as functions
of $\pac$ and $\phi$ once the acoustic field is known. The following
two extreme cases deserve special consideration:
\begin{itemize}
\item for a standing wave, $\phi(\vvr)=\phi_0$, so that
$\pgrad(\vvr)=\sdsurd{\pac}{\ri}$ and $\psi_i(\vvr)=\phi_0$,
\item for a traveling wave, $\pac(\vvr)=P_0$ and $\phi(\vvr)=-\vv{k}.\vvr$
so that $\pgrad(\vvr) = k_i P_0$ and $\psi_i(\vvr)=\phi(\vvr)-\pi/2$.
\end{itemize}

\subsection{Bubble model}
The radial oscillations of a gas bubble in a liquid under the action
of the sound field can be described by the Rayleigh-Plesset (RP)
equation
\cite{rayleigh,neppirasnol51,hilgenbrennergrosslohse98,linstoreyszeri2002pg}:
\begin{equation}
  \label{rp}
  \begin{split}
  \lhsrayleighp &=
  \frac{1}{\rho}\biggl[p_g+\frac{R}{c_l}\frac{dp_g}{dt}-
    4\mu\frac{\dot{R}}{R} \\
    &\quad -\frac{2\sigma}{R}-(p_0+\pact(t)) \biggr],
\end{split}
\end{equation}
where $p_0$ is the hydrostatic pressure, $p_g(t)$ is the gas pressure,
$\rho$, $\mu$ and $c_l$ are the density, viscosity and sound speed of
the liquid, respectively, and $\sigma$ is the surface tension.  The
ambient radius of the bubble $\rz$ is the radius that would have the
gas the in absence of the sound field.

Time is non-dimensionalized by the angular frequency $\omega$, and in
order to obtain a formulation consistent with
Ref.~\onlinecite{hilgenbrennergrosslohse98}, we set
\begin{equation}
  \label{adimrt}
  p_0 + \pact(\vvr,t) = p_0(1-p\cos x),
\end{equation}
so that 
\begin{eqnarray}
  \label{defp}
  p &=& \pac(\vvr)/p_0 \\
  \label{defx}
  x &=& \omega t + \phi(\vvr) - \pi.
\end{eqnarray}

Using $x$ as the time-variable, and non-dimensionalizing pressure with $p_0$,
equation (\ref{rp}) can be written as:
\begin{equation}
  \begin{split}
    \label{rpadim}
    R R'' + \frac{3}{2} R'^2 &=
    \frac{\rres^2}{3}\biggl[p_g^*+\frac{R\omega}{c_l}\frac{dp_g^*}{dx}-
      \frac{4\mu\omega}{p_0}\frac{R'}{R}\\
      &\quad -\alpha_S\frac{R_0}{R} + p\cos x -1 \biggr],
  \end{split}
\end{equation}
where primed variables denotes $d/dx$, 
\begin{equation}
  \label{defrres}
  \rres = \omega^{-1}(3p_0/\rho)^{1/2}
\end{equation}
is the resonance radius, and
\begin{equation}
  \label{defalphas}
\alpha_S = 2\sigma/p_0\rz  
\end{equation}
is the dimensionless Laplace tension.

Several models can be used for the bubble internal pressure $p_g$,
\cite{prosperetti91, prosperettihao99, prospernato, storeyszeri2001,
  toegel2000, brenner2002}. As will be seen below, we are mainly
interested here in the expansion phase of the bubble, during which the
density of the gas in the bubble remains weak, so that the precise
choice of the thermal bubble interior's model is unimportant.
However, in order to assess the validity of the approximate
expressions developed hereafter, simulations will be performed by
using the Keller equation \cite{kellerkolod,prosperlezzi1}. The bubble
interior is modeled by using a thermal diffusion layer following
Ref.~\onlinecite{toegel2000}, neglecting water evaporation and
condensation through the bubble wall.  In the remaining part of the
paper, we will consider air bubbles in water ($\sigma=0.072$
N.m$^{-1}$, $p_0=101300$ Pa, $\rho = 1000$ kg.m$^{-3}$, $c_l=$ 1498
m.s$^{-1}$, $\mu=$ $10^{-3}$ Pa.s).
 
\subsection{The Bjerknes force}
The primary Bjerknes force acting on a bubble is defined as
\begin{equation}
  \label{defbjerknes}
  \fbun = -\left<V(t) \grad \pact \right>,
\end{equation}
where $V(t)$ is the instantaneous bubble volume. The average is
taken over one acoustic period, so that, using Eq.~(\ref{defpgrad}):
\begin{equation}
  \label{defbjerknesint}
  \fbun_i = -\pgrad (\vvr) \frac{1}{T}\int_0^{T} V(t)  \cos
  \left[\omega t + \psi_i(\vvr) \right]  \; dt .
\end{equation}
Using the dimensionless time $x$ defined by (\ref{defx}) and the
periodicity of $V$, the latter expression becomes
\begin{equation}
  \label{defbjerknesintx}
  \fbun_i = \pgrad (\vvr) \frac{1}{2\pi}\int_0^{2\pi} V(x)  \cos
  \left[x-\phi(\vvr) + \psi_i(\vvr) \right]  \; dx.
\end{equation}

The generic problem is therefore to obtain an approximate analytical
expression for the integral
\begin{equation}
  \label{defI}
  I = \frac{1}{2\pi}\int_0^{2\pi} V(x)  \cos  \left(x-x_0 \right)  \; dx,
\end{equation}
valid for any bubble dynamics, and for any value of $x_0$. The problem
can be easily solved for small-amplitude linear oscillations
\cite{leightonwalton}. Here, we focus on the case of inertial oscillations,
that is for any combination of parameters $(p,R_0)$ above the Blake
threshold. The special cases of standing waves and traveling waves
can be simply recovered by setting, respectively, $x_0=0$ and $x_0=\pi/2$.

\section{Approximate expressions}
\label{secapprox}
\subsection{Bubble radius}
The method used to obtain analytical formula for the bubble radius are
mainly inspired from the approach of
\citet{hilgenbrennergrosslohse98}. For self-consistency, we will
recall in this section the main lines of the method, and, where
convenient, specify the refinements obtained by our approach.

\begin{figure}[ht]
  \includegraphics[width=\linewidth]{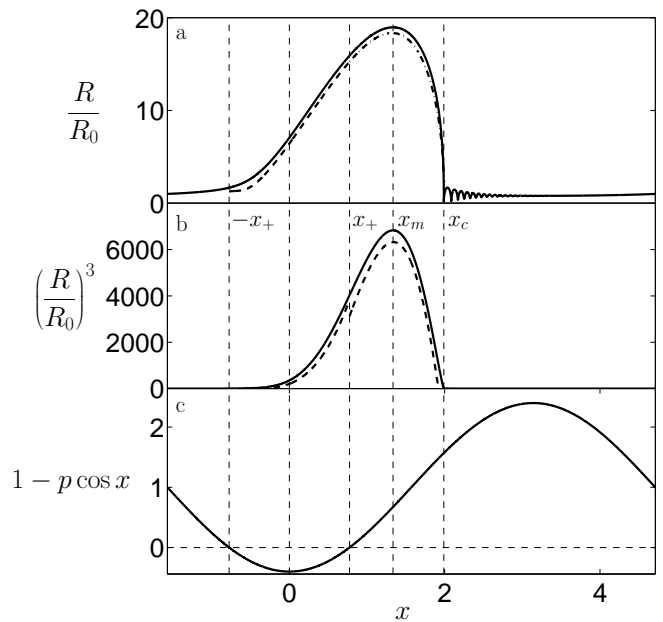}
  \caption{\label{figr} (a) Dimensionless bubble radius $R/R_0$; (b)
    Dimensionless bubble volume $(R/R_0)^3$; (c) Dimensionless driving
    pressure $1-p\cos x$. The case considered is a 3 $\mu$m air bubble
    in water and $p=1.4$. The times $-\xp$ and $\xp$ are the two
    instants of zero-crossing of the driving pressure, $x_m$ is the
    time of maximum expansion of the bubble, and $x_c$ the time of
    maximum compression. The dashed curve in (a) represents the
    approximate dynamics given by Eqs.  (\ref{solrmoins}),
    (\ref{solrplus}). The dashed line in (b) is the final
    approximation of the bubble volume (\ref{vmoins})-(\ref{vplus}).}
\end{figure}

Figure \ref{figr} displays the dimensionless bubble radius
(\ref{figr}a), bubble volume (\ref{figr}b), and driving pressure
(\ref{figr}c) in a typical case of inertial cavitation of ($f$ =
20~kHz, $R_0=$ 3 $\mu$m, and $p=$ 1.4). With the choice of the
dimensionless time-variable Eq.~(\ref{adimrt}), $x=0$ represents the time
of maximum tension of the liquid. We set
\begin{equation}
  \label{defxp}
  \xp = \acos{\frac{1}{p}},
\end{equation}
and we denote by $x_m$ the time of maximum expansion of the bubble,
and by $x_c$ the time of its maximum compression (see Fig.~\ref{figr}).

It is shown in Ref.~\onlinecite{hilgenbrennergrosslohse98} that,
during the expansion phase and most of the collapse phase, the
dominant terms in the right-hand-side of Rayleigh equation are the
driving term $p\cos x-1$ and also the surface tension term $\alpha_S
R_0/R$ for ambient radii just above the Blake threshold.  Following
Ref.~\onlinecite{hilgenbrennergrosslohse98}, we neglect the dependence
of the surface tension term in $R$ , and replace $\alpha_S R_0/R$ by
$\alpha_S/K(p)$, where $K(p)$ will be determined later. The
approximate Rayleigh equation becomes

\begin{equation}
  \label{rpapprox}
  R R'' + \frac{3}{2}R'^2 = \frac{\rres^2}{3} \left[
    p\cos x - \left(1 + \frac{\alpha_S}{K(p)}\right) \right].
\end{equation}
We set, for further use
\begin{equation}
  \label{defA}
  A = 1 + \frac{\alpha_S}{K(p)}.
\end{equation}
Besides, noting that
\begin{equation}
  \label{diff2r2}
  R R'' + R'^2 = 1/2 \frac{d^2(R^2)}{dx^2},
\end{equation}
the right-hand side of Rayleigh equation can be written in two
different forms:
\begin{eqnarray*}
  R R'' + \frac{3}{2}R'^2
  &=& \frac{1}{2} \frac{d^2(R^2)}{dx^2} + \frac{1}{2}\dot{R}^2 \\
  &=& \frac{3}{4} \frac{d^2(R^2)}{dx^2} - \frac{1}{2} R\ddot{R}.
\end{eqnarray*}

Numerical simulations show that $\dot{R}^2 \gg R\ddot{R}$ on the
interval $[-\xp, \xp]$, while $\dot{R}^2 \ll R\ddot{R}$ 
holds on the interval $[\xp, x_m]$ \cite{hilgenbrennergrosslohse98}.
Additionally, we found that the latter property still holds in fact
during almost all the collapse, except in its ultimate phase, where
the gas and acoustic terms become significant again. This could be
expected since the main part of the collapse is inertially driven and
that $\dot{R}$ becomes significant only when the liquid has acquired
enough kinetic energy.  We therefore obtain the following equations
for the bubble radius, over the interval $[-\xp,x_c]$:
\begin{eqnarray}
  \label{eqmoins}
  \frac{d^2(R^2)}{dx^2} &=& \frac{4}{9}\rres^2\left(p\cos x - A \right)
  \quad\text{on }[-\xp, \xp], \\
  \frac{d^2(R^2)}{dx^2} &=& \frac{2}{3}\rres^2\left(p\cos x - A \right)
  \quad\text{on }[\xp, x_c].
\end{eqnarray}
These equations are the same as the ones of
\citet{hilgenbrennergrosslohse98}, except that the validity of the
second is extended up to $x_c$.  The first equation can be solved with
the initial condition $R(-\xp) = \zeta R_0$,where $\zeta \simeq 1.6$
and $\dot{R}(-\xp)\simeq R(-\xp)$ \cite{hilgenbrennergrosslohse98}.
The second equation is solved by requiring continuity of $R(x)$ and
$R'(x)$ at $x=\xp$.  Integrating both equations twice, we obtain:
\begin{equation}
  \label{solrmoins}
  \begin{split}
    \rmoins^2(x) &= \frac{4}{9}\rres^2  \biggl[
    1 - p\cos x + p(x+\xp)\sin\xp  \\
    &\quad- \asd\left(x+\xp \right)^2 \biggr] \\
    &\qquad+ \zeta^2 R_0^2 \left[1+2\left(x+\xp \right) \right],
  \end{split}
\end{equation}
and
\begin{equation}
  \label{solrplus}
  \begin{split}
    \rplus^2(x) &= \frac{2}{3}\rres^2  \biggl[
      1 - p\cos x + p\left(\frac{x}{3}+\xp \right)\sin\xp \\
      &\quad- \frac{A}{2}\left(x^2+\xp^2+\frac{2}{3}\xp x \right)
    \biggr] \\
    &\qquad+ \zeta^2 R_0^2 \left[1+2\left(x+\xp \right) \right].     
  \end{split}
\end{equation}


The point $(x_m, \rmax)$ of maximum expansion is obtained by setting
$d(\rplus^2)/dx= 0$, so that $x_m$ is given in implicit form by
\begin{equation}
  \label{eqxm}
  \begin{split}
  & p\sin x_m -  x_m +\frac{1}{3}\left(p\sin\xp-\xp \right) \\
  &\quad   -\frac{\alpha_S}{K(p)}\left(x_m+\frac{1}{3}\xp \right) 
+ 3\zeta^2\left(\frac{R_0}{\rres} \right)^2 = 0,
  \end{split}
\end{equation}
%
%
%
and $\rmax$ reads
\begin{equation}
  \label{rmax}
  \begin{split}
    \rmax^2 &= R_0^2 f(p,x_m) \\
    &\quad  + \rres^2 \left[g(p,x_m)-\frac{2}{3}\frac{\alpha_S}{K(p)} h(p,x_m) \right],
  \end{split}
\end{equation}
where
\begin{eqnarray}
  \label{deff}
  f(p,x_m) &=& \zeta^2 \left[1+2\left(x_m+\xp \right) \right], \\
  \label{defg}
  \nonumber  g(p,x_m) &=& \frac{2}{3}
  \biggl[ 1 - p\cos x_m + p\left(\frac{x_m}{3}+\xp \right)\sin\xp \\
  & &\quad -\frac{1}{2}\left(x_m^2+\xp^2+\frac{2}{3}\xp x_m \right) \biggr], \\
  \label{defh}
  h(p,x_m) &=& \frac{1}{2}\left(x_m^2+\xp^2+\frac{2}{3}\xp x_m \right).
\end{eqnarray}

\begin{figure}[ht]
  \includegraphics[width=\linewidth]{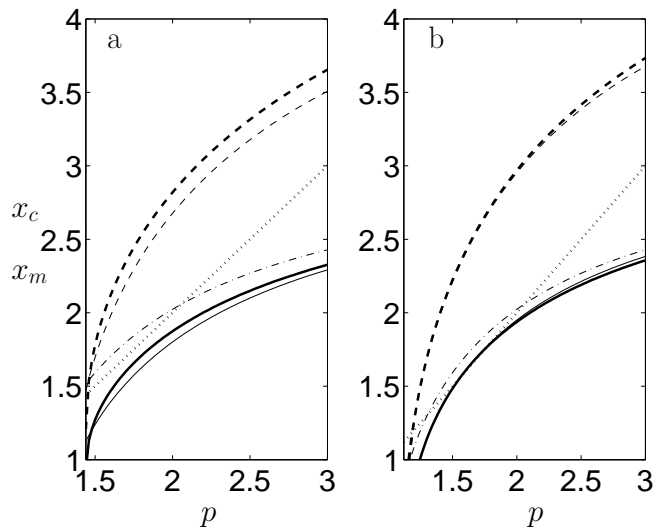}

  \caption{Thick solid line: $x_m$ calculated from numerical solutions of equation
    (\ref{rp}); Thin solid line: $x_{m_2}$ from Eq. (\ref{defxm2});
    Dash-dotted line: $x_{m_1}$ from Eq. (\ref{defxm1}) 
    (ref. \cite{louisnardgomez2003}); dotted line : $x_m = p$ 
    (ref. \cite{hilgenbrennergrosslohse98}).
    Thick dashed line: $x_c$ calculated from numerical solutions of equation
    (\ref{rp}); Thin dashed line: $x_c$ from Eq. (\ref{defyc}).
    The results are calculated for a bubble of ambient radius $R_0=$ 1
    $\mu$m (a), and $R_0=$ 3 $\mu$m (b). 
    \label{figxm}%
  }
\end{figure}

In order to obtain $x_m$, the implicit equation (\ref{eqxm}) should be
solved. To avoid this, Hilgenfeldt and co-workers
\cite{hilgenbrennergrosslohse98} developed this equation near $\pi/2$
at first order, neglecting on the one hand $\alpha_S/K(p)$, and also
$(R_0/\rres)^2$, which is appropriate for driving the bubble at low
frequencies. They obtain
\begin{equation}
  \label{defxm0}
  x_{m_0} = p + \frac{1}{3}\left(p\sin\xp-\xp \right),
\end{equation}
which can be further simplified as $x_m = p$, if $p$ is small enough.
Plugging the latter into Eqs. (\ref{rmax})-(\ref{defh}), they obtain an
expression of $\rmax$ which depends on $R_0$ only through the
$\alpha_S$ term in (\ref{rmax}). The expression of $K(p)$ is then
determined by using the fact, confirmed numerically, that the maximum
of the response curve $(\rmax/R_0)(R_0)$ is obtained for an
ambient radius $R_0^C$ very close to the Blake threshold
\begin{eqnarray}
  \label{condmax}
  \dsurd{}{R_0}\left(\frac{\rmax(p,R_0)}{R_0} \right) &=& 0 \nonumber\\
  \text{for}\quad R_0 = \rc &=& \frac{4\sqrt{3}}{9}\frac{\sigma}{p_0}\frac{1}{p-1}.
\end{eqnarray}
This scheme yields a good approximation for $\rmax$, which was the
main objective of Hilgenfeldt and co-workers
\cite{hilgenbrennergrosslohse98}, but the approximation (\ref{defxm0})
of $x_m$ yields a rather large error (see dotted line in Fig.
\ref{figxm} and \footnote{Curiously, it can be checked that although
  $x_m=p$ is a simplification of (\ref{defxm0}), it still yields
  better results in the whole range considered in Fig. \ref{figxm}.
  This is why we did not represent (\ref{defxm0}) on the latter.}).
Since the value of the integral (\ref{defI}) is found to be very
sensitive to the precise location of $x_m$, we seek a better
approximation.

We therefore revert to the original equations
(\ref{eqxm})-(\ref{defh}). The main difficulty lies in the presence of
the $\alpha_S$ term in (\ref{eqxm}), which makes rigorously $x_m$ a
function of both $p$ and $R_0$. Thus $\rmax$ depends not only on $R_0$
through $\alpha_S$ but also through $x_m$ in the expressions of $f$,
$g$ and $h$. The condition (\ref{condmax}) therefore becomes more
complex, and should be solved simultaneously with (\ref{eqxm}). We
initially followed this complex process, but finally found that a
better approximation of $x_m$ could be obtained by using a simple
trick. First, as was done in Ref.~\onlinecite{louisnardgomez2003}, we
neglect the $\alpha_S$ term in (\ref{eqxm}) and develop the latter
near $\pi/2$, but up to second order :
\begin{eqnarray}
  \label{defxm1}
  x_{m_1} &=& \frac{\pi}{2} -\frac{1}{p} + \frac{1}{p}\biggl\{ 1 + 2p
  \nonumber  \\
  & &\quad\times
  \biggl[x_{m_0} - \frac{\pi}{2} + 3\zeta^2\left(\frac{R_0}{\rres} \right)^2
  \biggr] 
  \biggr\}^{1/2}.
\end{eqnarray}
For low frequency driving, $\rz\ll\rres$, and   $x_{m_1}$
depends only slightly on $R_0$. We then plug (\ref{defxm1}) in Eqs
(\ref{rmax})-(\ref{defh}) and express the condition (\ref{condmax}),
neglecting $\sdsurd{x_m}{R_0}$, to obtain
\begin{equation}
  \label{defK1}
  K_1(p) = \frac{\xmun^2+\xp^2+\frac{2}{3}\xp \xmun}{g(p,\xmun)} 
  \frac{9}{4\sqrt{3}}(p-1).
\end{equation}
We now expand again (\ref{eqxm}) near $\pi/2$, but keeping the
$\alpha_S$ term, in which we set $K = K_1(p)$, to obtain
\begin{eqnarray}
  \label{defxm2}
  x_{m_2} &=& \frac{\pi}{2} -\frac{A_1}{p} + \frac{1}{p}\biggl\{ A_1^2
  + 2p \nonumber \\
  & &\quad\times
  \biggl[x_{m_0} - A_1\frac{\pi}{2} + \left(1-A_1\right)\frac{\xp}{3}+
  \nonumber \\
  & & \quad\quad 3\zeta^2\left(\frac{R_0}{\rres} \right)^2
    \biggr] 
  \biggr\}^{1/2},
\end{eqnarray}
%
%
where 
\begin{equation}
  \label{defA1}
  A_1 = 1+\frac{\alpha_S}{K_1(p)}.
\end{equation}
Setting $A_1 = 1$ in $x_{m_2}$, that is, neglecting the effect of
surface tension, the result of Ref.~\onlinecite{louisnardgomez2003},
Eq.~(\ref{defxm1}), is recovered.
 
Figure \ref{figxm} represents the variations of $x_m$ for a bubble of
ambient radius $R_0=$ 1 $\mu$m (fig.~\ref{figxm}.a) and $R_0=$ 3
$\mu$m (fig.~\ref{figxm}.b) in water. The thick solid lines are the
exact value obtained numerically, and the thin solid lines represent
$x_{m_2}$. The agreement is seen to be excellent, although a
noticeable difference can be seen for $R_0=$ 1 $\mu$m, which
originates from the over-simplification done when accounting for
surface tension by the simple term $\alpha_S/K(p)$ in
Eq.~(\ref{rpapprox}).  Also shown is the approximation $x_{m_1}$
(dash-dotted line), which does not take surface tension into account.
This clearly introduces an noticeable error on $x_m$, reasonably
corrected by Eq. (\ref{defxm2}). Finally, the approximation $x_m = p$
proposed in Ref.~\cite{hilgenbrennergrosslohse98} is displayed (dotted
line).

Finally, the approximation of $\rmax$ can then easily
be obtained by plugging an approximation of $x_{m}$ into
Eq.~(\ref{rmax}). This was done in
Ref.~\onlinecite{louisnardgomez2003} using $x_{m_1}$, and an excellent
agreement was found. The gain brought by using $x_{m_2}$ instead of
$x_{m_1}$ in (\ref{rmax}) remains unimportant, and for brevity, we do
not present the comparison between the analytical and numerical
expressions of $\rmax$ here.
\subsection{Bubble volume}
Approximations of the bubble volume could readily be obtained from the
approximations (\ref{solrmoins}),~(\ref{solrplus}) of the bubble
radius.  However, such expressions do not yield analytical expressions
of the integral (\ref{defI}) in closed form, and further
approximations are therefore required. First, we consider frequencies
low enough to have $\rres\gg R_0$, so that the $\zeta$ term can be
safely neglected in equations (\ref{solrmoins})-(\ref{solrplus}).

\subsubsection{Approximate expression on $[-\xp,\xp]$}

Numerical simulations demonstrate that $\rmoins$ is almost linear
between 0 and $\xp$ (Fig. \ref{figr}.a), which suggests that Eq.
(\ref{solrmoins}) is almost a perfect square in this interval.   We
then develop the $\cos$ term in (\ref{solrmoins}) near $x=0$ up to
second order, and write the result as
\begin{equation}
  \begin{split}
    \rmoins^2(x) &= \frac{4}{9}\rres^2 \biggl\{
    \frac{p-A}{2} \left[x + \frac{p\sin\xp-A\xp}{p-A} \right]^2 \\
    &\quad +  p\xp\sin\xp - \frac{A}{2}\xp^2 + 1-p  \\
    &\quad\quad - \frac{(p\sin\xp-A\xp)^2}{2(p-A)} \biggr\}
\end{split}
\end{equation}
For $\rmoins$ to be linear in $x$, the constant term in the bracket
must be negligible, so that $\rmoins$ can be simplified as  
\begin{equation}
  \label{rmoinsapprox}
  \rmoins(x) = \frac{2}{3}\rres \sqrt{\frac{p-A}{2}}
  \left(x - \frac{A\xp-p\sin\xp}{p-A} \right).
\end{equation}
The expression of the bubble volume on $[-\xp, \xp]$ therefore reads

\begin{equation}
  \label{vmoins}
  \vmoins(x) = 
  \left(\frac{2}{3}\rres \sqrt{\frac{p-A}{2}} \right)^3
  (x-x_1)^3,
\end{equation}
where 
\begin{equation}
  \label{defx1}
  x_1 = \frac{A\xp-p\sin\xp}{p-A},
\end{equation}
which allows to calculate integral (\ref{defI}) in closed form.

\subsubsection{Approximate expression on $[\xp,x_c]$}
Using equations (\ref{eqxm}) and (\ref{rmax}), it can be easily
checked that, setting $y=x-x_m$, the expression (\ref{solrplus}) of
$\rplus$ can be recast as:
\begin{equation}
  \label{solrplusbis}
  \rplus^2 = \rmax^2 + \frac{2}{3}\rres^2 L(y),
\end{equation}
where
\begin{eqnarray}
  \label{defL}
  L(y) &=& 2p\cos x_m\sin^2\frac{y}{2}-A\frac{y^2}{2} \nonumber \\
  & &\quad +  p\sin x_m \left(\sin y-y \right).
\end{eqnarray}
The bubble volume on $[\xp, x_c]$ becomes therefore
\begin{equation}
  \label{defvplus}
  \vplus = \rmax^3\left[1+\frac{2}{3}\left(\frac{\rres}{\rmax}
    \right)^2 L(y) \right]^{3/2},
\end{equation}
which unfortunately does not yield an explicit integration of
(\ref{defI}). Further progress can be done by noting that, from
Eq.~(\ref{rmax}), $\rres$ and $\rmax$ are of the same order of
magnitude, and that from (\ref{defL}), $L(y) = O(y^2)$ near $y=0$.
Equation (\ref{defvplus}) can therefore be approximated by
\begin{equation}
  \label{defvplusdev1}
  \vplus = \rmax^3\left[1+\left(\frac{\rres}{\rmax} \right)^2 L(y) + O(y^4)\right].
\end{equation}
Thus, to the same order of approximation, $L(y)$ can be replaced by
any equivalent expression up to order 4 in $y$, and the choice must be
directed by the ability of $\vplus\cos(x-x_0)$ to be integrable in
closed form.  We therefore choose to set $y^2/2 = \sin^2(y/2) +
O(y^4)$ and $\sin y-y = -1/6 \sin^3 y + O(y^5)$ in Eq. (\ref{defL})
to finally obtain:
\begin{eqnarray}
  \label{vplus}
  \vplus(x) &=&  \rmax^3 + \rmax\rres^2
  \biggl[2(p\cos x_m-A)\sin^2\frac{y}{2} \nonumber \\
  & &\quad -\frac{1}{6}p\sin x_m \sin^3 y  \biggr] + O(y^4),
\end{eqnarray}
which can now yield an explicit expression for integral (\ref{defI}).
 
It can further be noted that neglecting the $\sin^3 y$ term in the
square bracket, and setting $\sin^2\frac{y}{2} \simeq y^2/4$, $\vplus$
is found to be zero for
\begin{equation}
  \label{defyc}
  y_c = x_c-x_m = \frac{\rmax}{\rres}\left( \frac{2}{A-p\cos x_m}\right)^{1/2},
\end{equation}
which constitutes a simple approximation of the collapse-time. The
comparison between this expression and the exact instant of minimum
radius is visible in Fig. \ref{figxm} (dashed lines). Here again, an
excellent agreement is found, but deteriorates toward small bubble
radii.


\section{Bjerknes force}
\subsection{Analytical expression}
With the expressions of the bubble volume (\ref{vmoins})
and~(\ref{vplus}) at hand, the integral (\ref{defI}) can be calculated
in analytical form, keeping the contribution of the integrand only in
the  intervals $[0, \xp]$ and $[\xp, x_c]$, since $V$ can be
neglected in the other regions (see Fig. \ref{figr}.b). The integral
is thus the sum of the two contributions:
\begin{equation}
  \label{defIdecomp}
  I = \imoins + \iplus,
\end{equation}
where
\begin{eqnarray*}
  \label{defImoinsIplus}
  \imoins &= \displaystyle 
  \int_0^{\xp} \vmoins(x)  \cos  \left(x-x_0 \right)  \; dx\\,
  \iplus &= \displaystyle 
  \int_{\xp}^{x_c} \vplus(x)  \cos  \left(x-x_0 \right)  \; dx.
\end{eqnarray*}
%

Using the approximate expressions (\ref{vmoins}) and (\ref{vplus})  of
the bubble volume, integration yields
\begin{eqnarray}
  \label{Imoinsfinal}
    \imoins &=& \frac{8}{27}\rres^3 \left( \frac{p-A}{2} \right)^{3/2} \nonumber\\
    & & \times \biggl[ \dx(\dx^2-6)\sin(\xp-x_0) \nonumber\\
    & &\quad + 3(\dx^2-2)\cos(\xp-x_0) \nonumber\\
    & &\quad+ x_1(6-x_1^2)\sin x_0 + 3(2-x_1^2)\cos x_0
   \biggr ],
 \end{eqnarray}
with
\begin{equation*}
  \label{defdeltax}
  \dx = \xp-x_1.
\end{equation*}
The contribution $\iplus$ reads
\newcommand{\isincar}[1]{f_2(#1)}
\newcommand{\isincub}[1]{f_3(#1)}
\begin{eqnarray}
  \label{Iplusfinal}
    \iplus &=& \rmax^3\left[ \sin(x_c-x_0) - \sin(\xp-x_0) \right]\nonumber \\
    & & +  \rmax \rres^2 \biggl\{\frac{1}{4}(p\cos x_m-A)\nonumber\\
    & &\quad\times \bigl[\isincar{y_c}-\isincar{\yp} \bigr] \nonumber\\
    & & \quad- \frac{1}{192}p\sin x_m
    \left[\isincub{y_c}-\isincub{\yp} \right]
    \biggr\},    
  \end{eqnarray}
where
\begin{eqnarray*}
  \label{defIsin2}
  \isincar{y} &=& 4\sin(y-y_0)-\sin(2y-y_0)-2y\cos y_0, \\
  \isincub{y} &=& 2\cos(2y+y_0)+\cos(4y-y_0) \\
  &\quad& +  12y\sin y_0-6\cos(2y-y_0),
\end{eqnarray*}
and
\begin{equation*}
  \label{defy}
  y_0 = x_0 - x_m, \quad \yp = \xp-x_m.
\end{equation*}

\begin{figure}[ht]
  \includegraphics[width=0.9\linewidth]{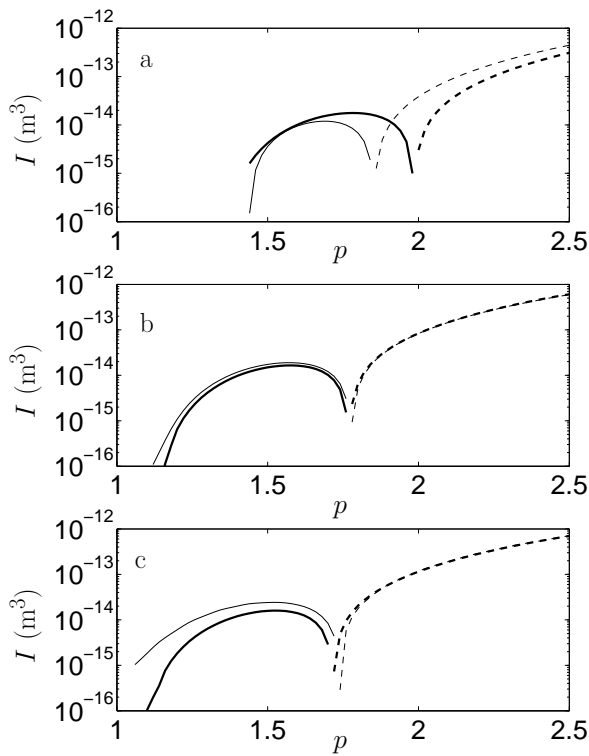}
  \caption{Thick lines: value of $I$ predicted by approximation
    (\ref{defIdecomp})-(\ref{Iplusfinal}), for $x_0=0$ (standing
    wave), for an air bubble in water, of radius $R_0=1$ $\mu$m (a), 3
    $\mu$m (b) and 6 $\mu$m (c). The solid parts of the curves
    correspond to $I>0$ and the dashed parts to $I<0$.  
    \label{figvcosx}%
  }
\end{figure}

The value of $I$ from (\ref{defIdecomp})-(\ref{Iplusfinal}) is
displayed in figure \ref{figvcosx} (thick lines) for $R_0=1$ $\mu$m
(Fig. \ref{figvcosx}a), 3 $\mu$m (Fig. \ref{figvcosx}b), and 6 $\mu$m
(Fig. \ref{figvcosx}c), in the case of a standing wave ($x_0=0$), for
drivings ranging from the Blake threshold to $p=2.5$. In order to get
a clear picture, $I$ is drawn in logarithmic scale, the solid part of
the curves representing a positive sign and the dashed part a negative
sign. The thin lines are the results obtained by solving (\ref{rp})
and calculating (\ref{defI}) numerically, for $f=20$ kHz. It is seen
that an excellent agreement is obtained, except for $R_0=1$ $\mu$m
(Fig~\ref{figvcosx}a).  Particularly, the point of inversion of the
Bjerknes force is shifted toward large drivings. This feature
originates from the errors induced on the values of $x_m$, $x_c$ (see
Fig.~\ref{figxm}) and $\rmax$ for small ambient radii, by replacing
the surface tension in the RP equation by $\alpha_S/K(p)$ in
(\ref{rpapprox}). It should be noticed that even the small errors
visible on the curves of Fig.~\ref{figxm}.a yields large differences
on the estimation of $I$. This could be expected since the phase
between $V$ and $\cos(x-x_0)$ crucially influences the value of 
integral $I$.

\subsection{Bjerknes force inversion threshold  in standing waves}
We consider the case of a standing wave $x_0=0$, and look for an
approximate locus in the parameter space where the Bjerknes force
changes sign. Summing equations (\ref{Imoinsfinal}) and
(\ref{Iplusfinal}), it is seen that integral (\ref{defI}) is zero for

\begin{figure}[ht]
  \includegraphics[width=\linewidth]{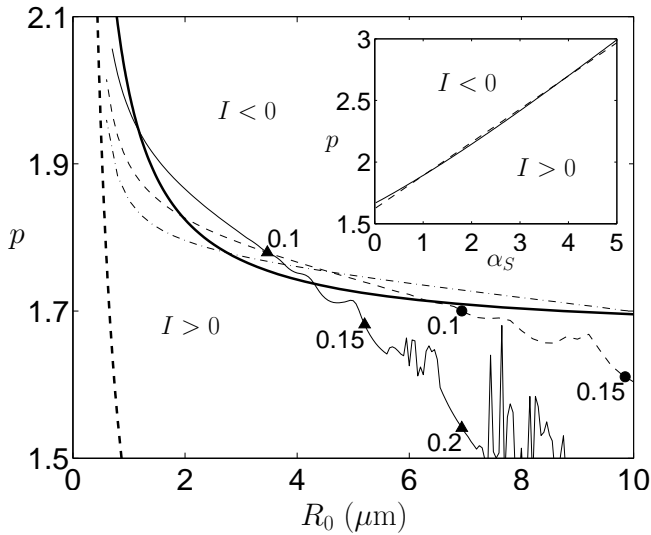}
  \caption{Threshold of Bjerknes force inversion in the $(R_0,p)$
    plane, for a bubble in water in ambient conditions ($\sigma=0.072$
    N.m$^{-1}$, $p_0=101300$ Pa, $\mu=$ $10^{-3}$ Pa.s). 
    The region $I>0$ corresponds to
    attraction by the pressure antinode, and $I<0$ to repulsion.
    The thin lines are calculated from numerical simulations of the
    RP equation. Thin solid line: $f=80$ kHz; dashed line: $f=40$ kHz;
    dash-dotted line:$f=20$ kHz. The labels on the curves indicate the
    ratio $R_0/\rres$ (triangles: $f=80$ kHz; filled circles: $f=40$
    kHz). Thick solid line: universal threshold calculated from
    approximate dynamics by solving  (\ref{eqannuleFbjXbis}). Thick
    dashed line: Blake threshold. The inset represent the
    solution of (\ref{eqannuleFbjXbis}) in the $(\alpha_S, p)$ plane.  
    \label{figinversionlocus}%
  }
\end{figure}

\newcommand{\rmxres}{\frac{\rmax}{\rres}}

\begin{equation}
  \label{fbjpol}
  a_3 X^3 + a_1 X + a_0 = 0
\end{equation}
where 
\begin{equation*}
  X = \rmxres,
\end{equation*}
and the coefficients $a_i$ depend on 
\begin{itemize}
\item $\xp$, which is just $\acos(1/p)$,
\item $x_1$, which from (\ref{defx1}) depends on $p$ and $\alpha_S$,
\item $x_c$, which from (\ref{defyc}) depends on $p$, $x_m$, $X$ and
  $\alpha_S$,
\item $x_m$, which from (\ref{defxm2}), only depends
  on $p$ and $\alpha_S$, for $R_0/\rres\ll 1$.
\end{itemize}
Furthermore, looking at Eq. (\ref{rmax}), for $R_0\ll\rres$,
$X = \rmax/\rres$ can be written as
\begin{equation}
  \label{rmaxrres}
  X = \left[g(p,x_m)-\frac{2}{3}\frac{\alpha_S}{K(p)} h(p,x_m) \right]^{1/2},
\end{equation}
and from Eqs. (\ref{deff})-(\ref{defh}) and
(\ref{defxm1})-(\ref{defA1}), still under the assumption
$R_0\ll\rres$, the terms $g$, $h$ in the above equation depend on
$R_0$ only through $\alpha_S$.  We conclude that, provided that
$R_0\ll\rres$, equation (\ref{fbjpol}) becomes frequency independent,
and can in fact be written in implicit form as
\begin{equation}
  \label{eqannuleFbjXbis}
  I(\alpha_S, p) = 0.
\end{equation}
This equation can easily be solved for $\alpha_S(p)$, in order to find
the approximate, frequency-independent threshold for inversion of the
Bjerknes force. The solution is presented in the inset of
Fig.~(\ref{figinversionlocus}). Below the curve, $I>0$, so that the
Bjerknes force attracts the bubble toward pressure antinodes, while it
becomes repulsive above.

From $\alpha_S = 2\sigma/(p_0 R_0)$, the inversion threshold can also
be plotted in the $(R_0,p)$ plane in the case of water at ambient
pressure ($\sigma=0.072$ N.m$^{-1}$, $p_0=101300$ Pa). The result is
displayed in Fig.  \ref{figinversionlocus} (thick solid line) and
compared to the exact inversion thresholds calculated from numerical
simulation for three driving frequencies 20 kHz (dash-dotted line), 40
kHz (dashed line), and 80 kHz (thin solid line). The labels on the two
latter curves represent the value of $R_0/\rres$. It is seen that the
above procedure yields a good estimation of the inversion threshold,
up to $R_0/\rres=0.1$, above which it starts to diverge from the exact
value. The reasons for this disagreement comes from the neglected
$R_0/\rres$ term in all expressions, and also from the fact that for
increasing frequency, the bubble rebounds become more important, so
that the bubble dynamics for $x>x_c$ also contributes to expression
(\ref{defI}). Besides, a cascade of period-doubling bifurcations and
chaos \cite{lautercra81prl,parlitz90,responsecurvesandmore} appear in
some cases (and are responsible for the noisy oscillations on the 80
kHz curve), so that the correct averaging of the Bjerknes force in
such cases should be carried out over more than a single acoustic
period. We did not pursue further this issue, since analytical
predictions for these bifurcations are out of the scope of the present
paper.

Marginally, it can be seen that the inversion threshold in the
$(\alpha_S,p)$ plane is almost linear, so that the following linear
fit (represented by a dashed line in the inset of
Fig.~\ref{figinversionlocus}) can be proposed for practical
applications:
\begin{equation}
  \label{fitseuilinversion}
  p = 0.269\;\alpha_S +  1.62.
\end{equation}
%
These results suggest that the inversion threshold is independent of
frequency, and of the properties of the gas and liquid other than
surface tension, as long as $R_0/\rres\ll 1$. This astonishing
result originates from the fact that the Bjerknes force mainly depends
on the expansion phase of the bubble, which, within the approximations
leading to Eq. (\ref{rpapprox}), is merely governed by the driving
pressure amplitude and surface tension. The reasonably good agreement
found in Figs. \ref{figxm}-\ref{figinversionlocus} partially supports
this analysis.

\begin{figure}[ht]
  \includegraphics[width=\linewidth]{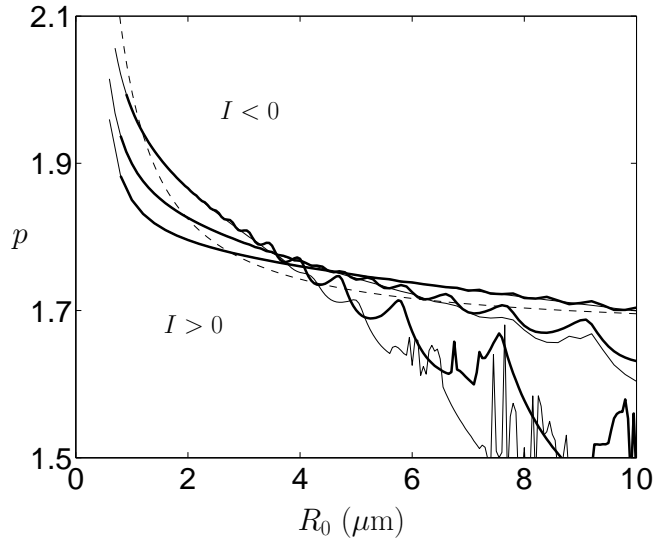}
  \caption{Same as Fig.~\ref{figinversionlocus}. The thin solid lines
    are the numerical curves of Fig.~\ref{figinversionlocus} ($f =$
    20, 40, 80 kHz).
    The thick solid lines are calculated in the same
    conditions, except that the gas behavior is considered isothermal.
    The thin  dashed line is the analytical threshold calculated from
    Eq.~(\ref{eqannuleFbjXbis}).%
    \label{figcomparethresholds}
  }
\end{figure}

In order to further investigate this issue, we first recalculated the
three inversion thresholds of Fig.~\ref{figinversionlocus} ($f=$ 20,
40, 80 kHz), replacing the thermal model of
Ref.~\onlinecite{toegel2000} by an isothermal behavior for the bubble
interior. Figure~\ref{figcomparethresholds} displays the results
obtained (thick solid lines) and recalls the thresholds calculated in
Fig.~\ref{figinversionlocus} (thin solid lines). It can be seen that
the thresholds slightly diverge for increasing $R_0$, but remain
almost indistinguishable for $R_0/\rres < 0.15$
We also repeated the calculations with the thermal model of
Ref.~\onlinecite{toegel2000}, but for argon bubbles (not shown), and found a
negligible deviation from the air curves.  We therefore conclude that
the detailed bubble interior has a very weak influence on the
expansion phase, at least for low enough values of $R_0/\rres$, so
that Eq.~(\ref{fitseuilinversion}) indeed constitutes a
gas-independent law, within its range of validity.

\begin{figure}[ht]
  \includegraphics[width=\linewidth]{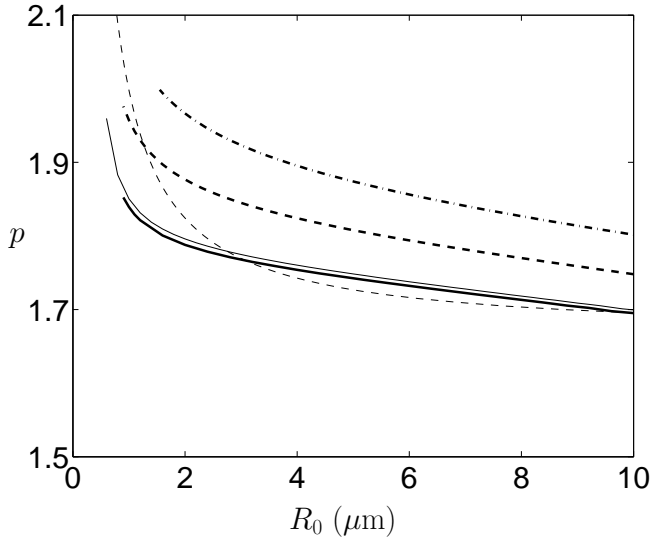}
  \caption{Same as Fig.~\ref{figinversionlocus}, only for $f=20$ kHz, and for
    different liquid viscosities. Thin solid line: water (same as dash-dotted
    line of Fig.~\ref{figinversionlocus}); Thick solid line: $\mu =
    0.1\,\mu_{\text{water}}$; Thick dashed line: $\mu = 10\,\mu_{\text{water}}$;
    Thick dash-dotted line $\mu = 20\,\mu_{\text{water}}$. The thin  dashed
    line is the analytical threshold calculated from Eq.~(\ref{eqannuleFbjXbis}). 
    \label{figcomparethresholdsvis}%
  }
\end{figure}

Another issue is the sensitivity of the results to the liquid
viscosity. The latter has been neglected in the analytical approach,
when approximating the RP equation~(\ref{rpadim}) by Eq.
(\ref{rpapprox}).  The good agreement found in
Fig.~\ref{figinversionlocus} between analytical and numerical results,
calculated for water at ambient temperature ($\mu=$ $10^{-3}$ Pa.s),
suggests that for such low values, viscosity indeed plays a minor role
during the bubble expansion.  One should however check whether it is
still the case for larger viscosities. We therefore repeated the
calculation of the inversion threshold for viscosities 10 and 20 times
larger than the one of water (Fig.  \ref{figcomparethresholdsvis},
thick dashed line and thick dash-dotted line). It is clearly seen that
the threshold increases noticeably with viscosity. Conversely, we also
checked that the result was unaffected by decreasing the viscosity
below the water's one, by computing the threshold for $\mu =
0.1\,\mu_{\text{water}}$ (thick solid line). This indicates
that viscous friction plays a non-negligible role in the bubble
expansion for viscosities above some critical value. As already
mentioned in Ref.~\onlinecite{hilgenbrennergrosslohse98}, increasing
viscosity decreases $\rmax$, and we also checked that it decreases
$x_m$ too, so that, strictly speaking, the Bjerknes force and its
inversion threshold are viscosity-dependent. Following our results,
this influence is negligible for  viscosities near or lower
than the water's one, but for slightly larger values, the
viscous term should be kept in the RP equation. 

\section{Discussion}

Important conclusions can be drawn from these results.  Figure
\ref{figinversionlocus} shows that the inversion thresholds for all
frequencies (thin lines) asymptotically merge with the Blake threshold
(thick dashed line) for small bubble radii, and in reasonable
agreement with the analytical approximation (thick solid line).
Thus, as the driving pressure reaches, say 1.8 bar, the range of
ambient radii of inertial bubbles attracted toward the antinode is
suddenly reduced, with an upper limit lower than 2 $\mu$m. This
explains why a well-defined bubble-free region can be observed around
the pressure antinode for high amplitude standing waves
\cite{parlitz99}. The range of attracted bubbles is however not void,
which suggests that the zone around the antinode could still be filled
with inertial bubbles, of ambient radii very close to the Blake
threshold, but too small to be visible. As noticed in Ref.
\cite{akhatov97une}, in a high amplitude standing wave, the Bjerknes
force acts as a sorter of inertial bubbles, leaving the smallest ones
approaching or even reaching the pressure antinodes.  The advantage of
the present analysis is that it yields, through Eq.
(\ref{eqannuleFbjXbis}), or its simpler form
(\ref{fitseuilinversion}), an explicit classification of the bubble
sizes as a function of the local acoustic pressure, parametrized by the
ratio~$\sigma/p_0$.

As the increasingly small bubbles approach the pressure antinode, they
may coalesce or quickly grow by rectified diffusion
\cite{louisnardgomez2003}. Increasing their size, they may again enter
the repulsion zone in the $(R_0,p)$ plane and move back again. This
picture is still complicated by the potential appearance of surface
instabilities.
Thus, the apparently void region observed around large pressure
antinodes may be in fact the locus of the complex evolution of very
small bubbles, of sizes close to the Blake threshold.

Finally, it is seen from the inset of Fig. \ref{figinversionlocus}
that decreasing $\alpha_S$ lowers the driving at which the pressure
antinode becomes repulsive. The dimensionless parameter $\alpha_S$ can
be varied experimentally by modifying the surface tension $\sigma$
(for example adding ionic salts or surfactants), or by changing the
static pressure $p_0$.  The present results suggest that, for
identical bubble ambient radii, the Bjerknes force would become
repulsive for lower drivings, when either decreasing $\sigma$ or
increasing $p_0$.  This should have an observable effect on the size
of the bubble-free region around the pressure antinode. However, it
should be noted that surface tension also plays a crucial role for
bubbles surface instabilities
\cite{plesset54,haoprosperetti99,brenner2002}, and also for rectified
diffusion \cite{louisnardgomez2003}, through the same dimensionless
parameter $\alpha_S$. Thus, changing $\alpha_S$ may also directly
influence these two processes, with probable consequences on the
bubble cloud behavior. The present result just demonstrates that
surface tension can influence the shape of the bubble cloud through
its direct effect on the bubble dynamics, and on the primary Bjerknes
force.

Figure \ref{figcomparethresholdsvis} also indicates that the size of
the bubble-free region around the pressure antinode would decrease
noticeably when increasing viscosity slightly above the one of water.
As mentioned in Ref.~\onlinecite{hilgenbrennergrosslohse98}, this may
be easily achieved experimentally by adding glycerin in water. Here
again, such a macroscopic effect is mediated by the sensitivity of the
bubble dynamics to the physical properties. To account analytically
for this dependence on viscosity, the viscous term should be kept in
the Rayleigh equation, which renders the approximation scheme more involved.
A generalization of our analytical results to this case may be
addressed in a future study.

Finally, it is highly probable that the same effect of surface tension
could be observed on the secondary Bjerknes force, as suggested by
numerical simulations \cite{akhatov97deux}. The extension of the
present analytical method to the latter effect is difficult, first
because the expression of the secondary Bjerknes force also involves
the bubbles velocities, which are much more sensitive to
approximations than the bubble radius itself, and secondly because the
dynamics equation of the two bubbles must be coupled by a radiation
term.

\bibliographystyle{apsrev}



\end{document}